# On the lightweight authenticated semi-quantum key distribution protocol without Trojan horse attack

**Jun Gu [1], Tzonelih Hwang[*]**

*Department of Computer Science and Information Engineering, National Cheng Kung University, No. 1, University Rd., Tainan City, 70101, Taiwan, R.O.C.*

[1] isgujun@163.com

[*]**Responsible for correspondence:**
Tzonelih Hwang
Distinguished Professor
Department of Computer Science and Information Engineering,
National Cheng Kung University,
No. 1, University Rd.,
Tainan City, 70101, Taiwan, R.O.C.
Email: hwangtl@ismail.csie.ncku.edu.tw
TEL: +886-6-2757575 ext. 62524

# Abstract

Recently, Tsai et al. (Laser Phys. Lett. 17, 075202, 2020) proposed a lightweight authenticated semi-quantum key distribution protocol for a quantum participant to share a secret key with a classical participant. However, this study points out that an attacker can use a modification attack to make both participants share a wrong key without being detected. To avoid this problem, an improvement is proposed here.

## 1. Introduction

Quantum key distribution (QKD) protocol allows the involved participants to share an unconditionally secure key. The first QKD protocol was proposed by Bennet et al. [1] in 1984. Afterward, various types of QKD protocols [2-4] have been proposed. However, most of these QKD protocols have two difficulties in implementation. First, these QKD protocols need all the participants to have heavy quantum capabilities, such as quantum joint operation, quantum register, and so on. Hence, a participant with only limited quantum capabilities is hard to involve in them. Second, an assumed authenticated channel is the prerequisite for running these protocols. That is, all the above QKD protocols adopt an ideal classical channel where the transmitted information cannot be modified and the identities of the communicating parties cannot be impersonated.

To solve the first problem, the semi-quantum key distribution (SQKD) [5] is proposed. That is, with SQKD, a classical participant who just has restricted quantum capacities can share a secret key with a quantum participant who has unlimited quantum capacities. The first SQKD protocol was proposed by Boyer et al. [5] in 2007. Subsequently, several SQKD protocols [6-9] have been proposed. In these protocols, the classical participants are restricted to have parts of the following quantum operations: (1) prepare qubits in Z-basis $\{|0\rangle, |1\rangle\}$, (2) measure qubits with the Z-basis, (3) reorder the qubits via different quantum delay lines, and (4) send or reflect the qubits.

For the second problem, the authenticated quantum key distribution (AQKD) [10] protocol is proposed. Different from the QKD needing an authenticated classical channel, the AQKD uses several pre-shared master keys to help the participants share a secret session key instead. Several AQKD protocols [8, 11, 12] have been proposed in recent years.

Recently, Tsai et al. proposed a lightweight authenticated semi-quantum key distribution (ASQKD) protocol without the Trojan horse attack [13]. They claimed that their protocol can help the involved participants use two pre-shared master keys to share a session key. However, this study shows that Tsai et al.'s ASQKD protocol suffers from a modification attack. That is, an attacker can make both participants share a wrong key without being detected. To solve this problem, a solution is proposed.

The rest of this paper is organized as follows. Section 2 briefly reviews Tsai et al.'s ASQKD protocol. Section 3 shows the modification attack on Tsai et al.'s protocol and then proposes an improvement to avoid the attack. At last, a conclusion is given in Section 4.

## 2. A brief review of Tsai et al.'s ASQKD

In Tsai et al.'s ASQKD protocol [13], a Bell state $|\Phi^+\rangle = \frac{1}{\sqrt{2}}(|00\rangle + |11\rangle)$ and two unitary operations $I = |0\rangle\langle 0| + |1\rangle\langle 1|$ and $H = \frac{1}{\sqrt{2}}(|0\rangle\langle 0| + |0\rangle\langle 1| + |1\rangle\langle 0| - |1\rangle\langle 1|)$ are used. Assume that Bob is a classical participant who just can measure particles with Z-basis and performs $I$ or $H$ on the particles. Alice is a quantum participant with unrestricted quantum capacities. Moreover, Alice and Bob pre-share two $2n$-bit master keys $K_1=\{k_1^1,k_1^2,\cdots,k_1^{2n}\}$ and $K_2=\{k_2^1,k_2^2,\cdots,k_2^{2n}\}$. Then Tsai et al.'s ASQKD protocol can be described as follows:

**Step 1**: Alice generates $2n$ Bell states $B=\{(q_A^1,q_B^1),(q_A^2,q_B^2),\cdots,(q_A^{2n},q_B^{2n})\}$ in $|\Phi^+\rangle$ and picks out all the first particles and all the second particles in $B$ to form the ordered particle sequences $S_A=\{q_A^1,q_A^2,\cdots,q_A^{2n}\}$ and $S_B=\{q_B^1,q_B^2,\cdots,q_B^{2n}\}$, respectively. Subsequently, she performs $I$ or $H$ on $S_A$ according to $K_1$. That is, if $k_1^i=0$ ($1\leq i\leq 2n$), she performs $I$ on $q_A^i$. Otherwise, $H$ is performed on $q_A^i$. Then, she sends $S_B$ to Bob.

**Step 2**: For the particles received, Bob performs $I$ or $H$ on them the same as Alice did in Step 1. Then he measures them with Z-basis $\{|0\rangle,|1\rangle\}$ to obtain a measurement result sequence $M_B=\{m_B^1,m_B^2\cdots,m_B^{2n}\}$ where $m_B^i$ is the measurement result of $q_B^i$. Bob sends a classical message to notify Alice that he has accomplished his operations.

**Step 3**: Similar to Bob, Alice measures $S_A$ with Z-basis to obtain a measurement result sequence $M_A=\{m_A^1,m_A^2\cdots,m_A^{2n}\}$.

**Step 4:** According to $K_2$, Alice (Bob) divides $M_A$ ($M_B$) into two parts, the raw key part $RK_A$ ($RK_B$) and the checking part $C_A$ ($C_B$). That is, if $k_2^i=0$ ($1\leq i\leq 2n$), $m_A^i$ ($m_B^i$) will belong to $RK_A$ ($RK_B$). Otherwise, $m_A^i$ ($m_B^i$) belongs to $C_A$ ($C_B$). Subsequently, Alice (Bob) divides $C_A$ ($C_B$) into two parts $C_A^O$ ($C_B^O$) and $C_A^E$ ($C_B^E$) where $C_A^O$ ($C_B^O$) consists of the values in the odd positions of $C_A$ ($C_B$) and $C_A^E$ ($C_B^E$) consists of the values in the even positions of $C_A$ ($C_B$). Then, Alice (Bob) sends $C_A^E$ ($C_B^O$) to Bob (Alice).

**Step 5:** Alice (Bob) checks whether the received $C_B^O$ ($C_A^E$) is equal to $C_A^O$ ($C_B^E$) or not. If the error rate exceeds a predetermined value, this protocol will be aborted. Otherwise, Alice and Bob extract the session key $SK$ by performing privacy amplification [13] on the $RK_A$ and $RK_B$, respectively.

## 3. Modification attack and counterattack on Tsai et al.'s protocol

Tsai et al. claimed that the proposed ASQKD can ensure that the final shared key is correct and secure. However, this section shows that Tsai et al.'s protocol suffers from a modification attack. That is, an attacker Eve can use a single photon operation to make the shared session key inconsistent between both participants. Besides, to avoid this loophole, a simple solution is proposed here.

### 3.1. The modification attack on Tsai et al.'s ASQKD protocol

In Step 4 and Step 5, Alice and Bob directly use parts of the measurement results of the shared Bell states to detect the attacker. If an attacker Eve can make all the measurement results obtained by Bob are flipped and then flips all the announced classical check bits, the shared key will be inconsistent and this attack will not be detected.

For example, in Step 1, Eve intercepts all the particles transmitted from Alice to Bob. Subsequently, she performs a unitary operation $i\sigma_y = |0\rangle\langle 0| + |1\rangle\langle 1|$ on each qubit where $i\sigma_y$ can transform the initial state $|\Phi^+\rangle = \frac{1}{\sqrt{2}}(|00\rangle + |11\rangle) = \frac{1}{\sqrt{2}}(|++\rangle + |--\rangle)$ into $|\Psi^-\rangle = \frac{1}{\sqrt{2}}(|01\rangle - |10\rangle) = \frac{1}{\sqrt{2}}(|-+\rangle - |+-\rangle)$. This means

that whether Alice and Bob perform $H$ or not, their measurement results always are opposite. To facilitate the discussion later, assume that Alice's (Bob's) measurement results $M_A = 0_1 0_2 1_3 1_4$ ($M_B = 1_1 1_2 0_3 0_4$) and the first two bits are the check bits. Hence, in Step 4, Alice (Bob) sends the check bit $C_A^E = 0_2$ ($C_B^O = 1_1$) to Bob (Alice) via a classical channel without authentication. Subsequently, Eve directly flips all the transmitted classical bits. Then, Alice and Bob will receive the flipped check bits $C_B^{O'} = 0_1$ and $C_A^{E'} = 1_1$, respectively. In Step 5, during the attacker detection process, Alice and Bob find $C_B^{O'} = 0_1 = C_A^O$ and $C_A^{E'} = 1_1 = C_B^E$, respectively. Hence, both of them think there is no attacker and then use the remaining bits ($RK_A = 1_3 1_4$, $RK_B = 0_3 0_4$) as the shared raw key. After privacy amplification, Alice and Bob will extract two inconsistent session keys which cannot be used for further communication.

**3.2. A solution to avoid modification attack on Tsai et al.'s ASQKD protocol**

As mentioned above, because Alice and Bob directly use the measurement results to check the attack, Eve can simply flip all the transmitted qubits and classical bits to modify the shared key without being detected. Hence, if Bob and Alice perform a hash function [14] on the check bits and announce the hash function results instead, then Eve cannot directly flip all the classical bits to pass the attacker detection anymore. This problem is solved. The improved protocol can be described as follows.

**Step 1'-3'** are the same as **Step 1-3** in Section 2.

**Step 4'**: Similar to Step 4, Alice (Bob) obtains $\{RK_A, C_A^O, C_A^E\}$ ($\{RK_B, C_B^O, C_B^E\}$) from $M_A$ ($M_B$). Then, Alice (Bob) performs a pre-shared hash function on $C_A^E$ ($C_B^O$) to obtain $h(C_A^E)$ ($h(C_B^O)$). Finally, Alice (Bob) sends $h(C_A^E)$

($h(C_B^O)$) to Bob (Alice).

**Step 5'** After Alice (Bob) receives $h(C_B^O)$ ($h(C_A^E)$), she (he) performs the hash function on $C_A^O$ ($C_B^E$) to obtain $h(C_A^O)$ ($h(C_B^E)$) and checks whether the received $h(C_B^O)$ ($h(C_A^E)$) is equal to $h(C_A^O)$ ($h(C_B^E)$) or not. If they are not equal, this protocol will be aborted. Otherwise, Alice and Bob extract the session key $SK$ by performing privacy amplification on the $RK_A$ and $RK_B$, respectively.

## 4. Conclusions

This paper shows that Tsai et al.'s ASQKD protocol suffers from a modification attack. With this attack, an attacker can make the shared session key between the involved participants inconsistent. To solve this problem, an improvement without needing the involved classical participant to have any extra quantum capacities is proposed.

## Acknowledgment

We would like to thank the Ministry of Science and Technology of the Republic of China, Taiwan for partially supporting this research in finance under the Contract No. MOST 109-2221-E-006-168-; No. MOST 108-2221-E-006-107-.